\documentclass[twocolumn,superscriptaddress,aps,prl]{revtex4-1}

\usepackage{graphicx}
\usepackage{bm}
\usepackage{color}
\usepackage{epstopdf}
\usepackage{amsmath}
\usepackage{amssymb}
\usepackage{epstopdf}

\usepackage[urlcolor=blue,colorlinks=true,citecolor=blue,linkcolor=blue,pdfstartview={FitH},bookmarks=false]{hyperref}

\newcommand{\Imag}{\text{Im}}
\graphicspath{{fig/}{./fig/}{.}}

\begin{document}

\title{Quantum engineering of Majorana quasiparticles in one-dimensional optical lattices}

\author{Andrzej Ptok}
\email[e-mail: ]{aptok@mmj.pl}
\affiliation{Institute of Physics, M.\ Curie-Sk\l{}odowska University, \\ ul.\ Radziszewskiego 10, PL-20031 Lublin, Poland}
\affiliation{Institute of Nuclear Physics, Polish Academy of Sciences, \\ ul. E. Radzikowskiego 152, PL-31342 Krak\'{o}w, Poland}

\author{Agnieszka Cichy}
\email[e-mail: ]{agnieszkakujawa2311@gmail.com}
\affiliation{Institut f\"{u}r Physik, Johannes Gutenberg-Universit\"{a}t Mainz, \\
Staudingerweg 9, D-55099 Mainz, Germany}

\author{Tadeusz Doma\'{n}ski}
\email[e-mail: ]{doman@kft.umcs.lublin.pl}
\affiliation{Institute of Physics, M.\ Curie-Sk\l{}odowska University, \\ ul.\ Radziszewskiego 10, PL-20031 Lublin,  Poland}

\date{\today}

\begin{abstract}
We propose a feasible way of engineering Majorana-type quasiparticles in ultracold fermionic gases on a one-dimensional (1D) optical lattice. For this purpose, imbalanced ultracold atoms interacting by the spin-orbit coupling should be hybridized with a three-dimensional Bose-Einstein condensate (BEC) molecular cloud. By constraining the profile of an internal defect potential we show that the Majorana-type excitations can be created or annihilated. This process is modelled within the Bogoliubov-de Gennes approach. This study is relevant also to nanoscopic 1D superconductors where modification of the internal defect potential can be obtained by electrostatic means.
\end{abstract}


\maketitle


\paragraph*{Introduction.}

--- Ultracold quantum gases provide a unique opportunity for {\it quantum simulations} of interacting many-body systems~\cite{bloch.dalibard.12}.
Tremendous progress in experimental techniques in the last years has allowed to control all important ingredients of such simulations, giving an insight into physical mechanisms that eluded understanding in conventional, ``natural'' condensed matter setups.
In particular, both the depth of the periodic trapping potential and the lattice geometry can be controlled, offering a variety of opportunities for research. 
Experiments in which fermionic or bosonic gases are loaded into the optical lattices have been carried out~\cite{bloch.dalibard.08}. 
There has been significant experimental progress in the engineering of artificial gauge fields, the spin-orbit (SO) couplings or simulating non-Abelian fields~\cite{lin.compton.09}.

\begin{figure}[!b]
\centering
\includegraphics[width=\linewidth]{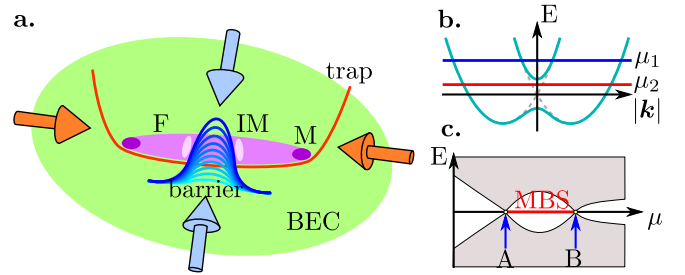}
\caption{
(a) An ultracold fermionic gas loaded on a 1D lattice (F) coupled to a molecular BEC in a trap (the red line).
The Majorana modes (M) are created on the edges of the wire.
Placing inside the trap an additional potential barrier (blue lines), one can effectively split the fermionic chain, creating additional two edges, which causes the induction of extra Majorana states (IM).
(b) Dispersion relation for a 1D population-imbalanced chain with spin-orbit interaction. 
The BEC medium induces pairing between opposite momenta and spin states, which creates a gap of size $\Delta$. 
There is the topologically (non)trivial phase for a Fermi level $\mu_{1}$ ($\mu_{2}$). 
(c) Exemplary energy spectrum $E$--$\mu$. The red line illustrates Majorana bound states (MBS), topologically protected inside the energy gap, occurring between A and B, which determine the boundary of the topologically non-trivial phase.          
\label{fig.schem}
}
\end{figure}

Recent studies of the topological matter, especially the topologically non-trivial  superconductivity, is motivated by realization of exotic quasiparticle excitations that resemble the Majorana fermions  (MFs)~\cite{wilczek.09, leijnse.flensberg.12}. In quantum field theory, MFs are particles that are their own antiparticles. In condensed matter, MF can be understood as a quasiparticle which is its ``own hole''. Moreover, the Majorana excitations have an exotic exchange statistics -- they are non-Abelian anyons~\cite{stern.10}, which makes them even more interesting. 
The prerequisites to observe such zero-energy Majorana modes in condensed matter systems are: a strong SO coupling, an external Zeeman magnetic field (population imbalance) and the existence of a gap in the energy spectrum ~\cite{mourik.zuo.12,nadjperge.drozdov.14,pawlak.kisiel.16,jiang.kitagawa.11}. 
The SO coupling is quite important from the point of view of real systems. 
It determines the electronic structure of atoms but also leads to such non-standard phenomena as the emergence of topological insulators. 
The SO coupling has been realized successfully in ultracold atomic gases setups. 
The  first experimental realization of the SO coupling has been performed in BEC using a two-photon Raman process~\cite{lin.jimenezgarcia.11}. 
As concerns the experimental realization of the SO coupling in an ultracold fermionic gas, it has been reported one year later~\cite{wang.yu.12}.    

So far, various experimental methods for detecting Majorana quasiparticles have used either semiconducting nanowires proximitized to superconductors~\cite{mourik.zuo.12,nadjperge.drozdov.14,pawlak.kisiel.16,ruby.15}, vortices in $p$-wave superconductors~\cite{sun.16}, or some lithographically designed nanostructures~\cite{suominen.17}. Finding feasible means for manipulating Majorana quasiparticles in such systems is still quite a big challenge. We address such issue for a nonuniform 1D system, where an additional pair of Majorana modes can be controllably created by designing a proper internal scattering potential.


\newpage

\paragraph*{Setup.} 

--- 
Our study is inspired by the proposal of L.~Jiang {\it et al.}~\cite{jiang.kitagawa.11} to investigate a trapped spin-imbalanced fermionic gas on a 1D optical lattice. This fermionic system is coupled to a 3D molecular BEC cloud (Fig.~\ref{fig.schem}.a), which provides the on-site {\it s-wave} pairing of atoms. By applying the spin-orbit and the Zeeman interactions (via a synthetic magnetic field), one can induce the {\it p-wave} pairing between the identical spin atoms from neighbouring sites. Such topologically non-trivial superconducting state is manifested by the zero-energy MBS appearing at the trap edges (practically, near the Fermi radius). We have checked that optimal conditions for MBS occur when a profile of the trapping potential is steep enough (see red line in Fig.~\ref{fig.schem}.a) as compared to the usual harmonic potential, because otherwise their spatial extent becomes rather fuzzy.

The main objective of our study here is to investigate how the energy spectrum of a 1D fermionic system (in particular revealing MFs) can be controllably affected by an additional potential, sketched by the blue lines in Fig.~\ref{fig.schem}.a. This quantum defect can be regarded as an internal boundary, which could induce a new pair of Majorana quasiparticles~\cite{maska.gorczyca.17} that  could be used e.g. for quantum computation~\cite{zhou.wu.14,spanslatt.ardonne.17,aliceaoreg.11}. Usually the easiest way to manipulate MFs is to use a T-shape geometry, where the Majorana edge states can be exchanged by electric means (using external gate potentials)~\cite{aliceaoreg.11}. Such processes would be useful for logical operations on quantum bits (qubit) made of the Majorana states~\cite{sato.takahashi.10,hyart.vanheck.13}.


\paragraph*{Microscopic model.} 

--- 
The 1D atomic Fermi chain on the optical lattice and coupled to the 3D molecular BEC (Fig.~\ref{fig.schem}.a) can be described by the following Hamiltonian 
\begin{eqnarray}
\mathcal{H} &=& \mathcal{H}_{0}  + \mathcal{H}_{SO}  + \mathcal{H}_{BEC} + \mathcal{H}_{trap} + \mathcal{H}_{bar}.
\end{eqnarray}
Here, $\mathcal{H}_{0} = \sum_{i,j\sigma} \left( - t \delta_{\langle i,j \rangle} - \left( \mu + \sigma h \right) \delta_{ij} \right) c_{i\sigma}^{\dagger} c_{j\sigma}$
describes free fermionic atoms in the lattice which hop between the nearest-neighbour sites with the hopping amplitude $t$, $\sigma=\uparrow ,\downarrow$ is the spin index, $\mu$ -- the chemical potential and $h$ is a Zeeman field which originates from a population imbalance.  
The spin-orbit coupling can be expressed by $\mathcal{H}_{SO} = - i \lambda \sum_{ \langle i,j \rangle \sigma \sigma' } c_{i\sigma} \left( \hat{\sigma}_{y} \right)_{\sigma\sigma'} c_{j\sigma'}$, where $\hat{\sigma}_{y}$ is the Pauli {\it y}-matrix.
Coupling of the BEC with the fermionic chain leads to the proximity induced on-site pairing which can be effectively modelled as $\mathcal{H}_{BEC} = \sum_{i} \left( \Delta^{\dagger} c_{i\downarrow} c_{i\uparrow} + H.c. \right)$. 

Here, $\Delta$ plays a role of the effective gap induced in the fermionic chain by the BEC background. 
Formally, $\Delta \approx g \Xi$~\cite{jiang.kitagawa.11}, where $g$ denotes the coupling constant between the composite bosonic and fermionic mixures, whereas $\Xi$ corresponds to macroscopic occupation in the ground state by composite bosons in the BEC~\cite{holland.kokkelmans.01}.
$\mathcal{H}_{trap} = \sum_{i\sigma} V ( {\bm r}_{i} ) c_{i\sigma}^{\dagger} c_{i\sigma}$ describes trapping potential ($\tilde{V}$), whereas $\mathcal{H}_{bar} = \sum_{i\sigma} \Lambda ( {\bm r}_{i} , t ) c_{i\sigma}^{\dagger} c_{i\sigma}$ describes the potential barrier ($\tilde{\Lambda}$) inside the trap. 

The dispersion relation for a fermionic chain with strong SO coupling, in a Zeeman magnetic field is displayed in  Fig.~\ref{fig.schem}.b. For the topologically trivial phase, the degeneracy of Fermi level $\mu_{1}$ is fourfold, whereas in a topologically non-trivial phase $\mu_{2}$ has only two-crossing points. By varying  $\mu$ the system qualitatively changes from the topologically non-trivial to topologically trivial one (Fig.~\ref{fig.schem}.c). In the first case, i.e. in the region located to the left of point A and to the right of point B, there is the standard gaped spectrum of the energy. 
However, between the point A and B in Fig.~\ref{fig.schem}.c, we are dealing with a topologically non-trivial phase which is manifested by the topologically protected MBS inside the soft gap (red line in Fig.~\ref{fig.schem}.c). The condition for the occurrence of such a topological phase in a homogeneous system is given by $\sqrt{ \Delta^{2} + ( 2 t + \mu )^{2} } > h$, and thus depends on the chemical potential $\mu_{2}$ in Fig.~\ref{fig.schem}.b.


\paragraph*{Methodological details.} 

--- The quasiparticle spectrum of ultracold atoms described by the Hamiltonian $\mathcal{H}$ can be obtained from a diagonalization procedure based on the Bogoliubov-Valatin transformation $c_{i\sigma} = \sum_{n} \left( u_{in\sigma} \gamma_{n} - \sigma v_{in\sigma}^{\ast} \gamma_{n}^{\dagger} \right)$. As usually, the BCS coefficients $u_{in\sigma}$ and $v_{in\sigma}$ fulfill the Bogoliubov--de Gennes (BdG) equations $\mathcal{E}_{n} \Psi_{in} = \sum_{j} \mathbb{H}_{ij} \Psi_{jn}$, where $\Psi_{in} = \left( u_{in\uparrow} , u_{in\downarrow} , v_{in\downarrow} , v_{in\uparrow} \right)$ is a four-component spinor, and the matrix 
\begin{eqnarray}
\label{eq.hammatrix} \mathbb{H}_{ij} &=&
\left(
\begin{array}{cccc}
H_{ij\uparrow\uparrow} & H_{ij\uparrow\downarrow} & \Delta_{ij} & 0 \\ 
H_{ij\downarrow\uparrow} & H_{ij\downarrow\downarrow} & 0 & \Delta_{ij} \\
\Delta_{ij}^{\ast} & 0 & -H_{ij\downarrow\downarrow}^{\ast} & H_{ij\downarrow\uparrow}^{\ast} \\ 
0 & \Delta_{ij}^{\ast} & H_{ij\uparrow\downarrow}^{\ast} & -H_{ij\uparrow\uparrow}^{\ast}
\end{array} 
\right)
\end{eqnarray}
is defined as: 
$H_{ij\sigma\sigma'} = \left( - t \delta_{ \langle i,j \rangle } - ( \mu + \sigma h ) \delta_{ij} \right) \delta_{\sigma\sigma'} + H_{SO}^{\sigma\sigma'}$, with $\Delta_{ij} = \Delta \delta_{ij}$. We introduce the following spin-orbit terms $H_{SO}^{\uparrow\downarrow} = \lambda ( \delta_{ i+1,j } - \delta_{ i-1,j } )$ and $H_{SO}^{\downarrow\uparrow} = ( H_{SO}^{\uparrow\downarrow} )^{\ast}$.

MBSs exist only in the non-trivial superconducting phase and correspond to the zero-energy modes $E_{n}=0$. For observing the MBS in the system, the following quantities can be considered:
{\it (i)} the local density of states (LDOS)~\cite{domanski.11},
{\it (ii)} the density of Majorana quasiparticles $\mathcal{P}_{M}$~\cite{sticlet.bena.12},
and
{\it (iii)} the topological quantum number $\mathcal{Q}$~\cite{akhmerov.dahlhaus.11}.
Below, we briefly discuss each of them.

The local density of states of fermionic atoms in a given site $i$ is defined as $\rho ( i , \omega ) = -1 / \pi \sum_{\sigma} \Imag G_{i\sigma,11} ( \omega + i 0^{+} )$, where $G_{i\sigma} ( \omega ) = \left( \omega - \mathbb{H} \right)^{-1}$ is the single particle Green's function with the matrix $\mathbb{H}$  given in Eq.~(\ref{eq.hammatrix}). Using the Bogolubov-Valatin transformation, one can find~\cite{matsui.sato.03,okazaki.ito.14}
\begin{eqnarray}
\rho ( i , \omega ) = \sum_{n,\sigma} \left[ | u_{in\sigma} |^{2} \delta \left( \omega - \mathcal{E}_{n} \right) + | v_{in\sigma} |^{2} \delta \left( \omega + \mathcal{E}_{n} \right) \right] , 
\end{eqnarray}
where $\delta ( \omega )$ is the Dirac delta function and $\mathcal{E}_{n}$ come from solving the BdG equations. 
The density of Majorana quasiparticles $\mathcal{P}_{M}$ is characterized by the off-diagonal spectral function at zero energy~\cite{sticlet.bena.12}
\begin{eqnarray}
\mathcal{P}_{M} ( i ) = \sum_{n} | u_{in\downarrow} v_{in\downarrow}^{\ast} -  u_{in\uparrow} v_{in\uparrow}^{\ast} | \delta ( E_{n} ) .
\label{density_P}
\end{eqnarray}
The density (\ref{density_P}) provides information about the spatial extent of MBS  in the system, which can be helpful for investigating non-locality of these quasiparticles.

Another important physical quantity is the topological number $\mathcal{Q} = ( - 1 )^{m}$, which tells about the fermion parity $m$ of the superconducting ground state~\cite{kitaev.01}. 
This quantity is helpful for the identification of the topological property of described system. 
Formally, $\mathcal{Q}$ can be appointed from the scattering matrix $\mathcal{S}$ which describes a relation between incoming and outgoing wave amplitudes at the Fermi level~\cite{akhmerov.dahlhaus.11}.
The $\mathcal{S}$ matrix is built by the blocks of the reflection $R$ and transmission $T$ matrices at the two ends of the system. 
In this case, $\mathcal{Q} = \mbox{sgn} \det ( R )$, and the sign denotes the possibility of the occurrence of the MBS only when $\mathcal{Q} = -1$.
Here, $\mathcal{Q}$ has been determined from the sign of the scattering matrix, within the BdG approach, following the numerical procedure discussed in ~\cite{zhang.nori.16}.
Formally, $\mathcal{Q}$ can be appointed from a Pfaffian of $R$ and can be regarded as the spin Chern number in a case of the $\mathbb{Z}_{2}$ topological phase~\cite{kane.05,fulga.11}.
Several methods for numerical determination of the topological quantum number have been proposed in the literature~\cite{moore.07,qi.08,roy.09}.  Of particular interest is for us the odd fermion parity, $\mathcal{Q} = -1$, refering to the topologically nontrivial superconducting phase~\cite{hasan.kane.10,qi.zhang.11}.


\paragraph*{Computational details.}

--- Our calculations have been done for a 1D lattice comprising $N = 600$ sites. We fixed the concentration of atoms $n \simeq 0.16$ and focused on the ground state ($T = 0$), assuming the model parameters $h = 0.3 t$, the SO coupling $\lambda = 0.15 t$ and the gap $\Delta = 0.2 t$.
This choice of parameters, in a case of the barrier potential absence, fulfills the condition for realisation of the Majorana quasiparticles in the system. To solve numerically the BdG equations, we have replaced the Dirac delta function by a narrow Lorentzian $\delta (\omega) = \zeta / [ \pi ( \omega^{2} + \zeta^{2} ) ]$ with a broadening $\zeta =0.002$ for LDOS and 10$^{-12}$ for $\mathcal{P}_{M}$.

We have investigated the trapping potential $\tilde{V}$ in the form of: {\it (i)} the parabolic trap $V ( {\bm r}_{i} ) = V_{0} ( r_{i} - N/2 )^{2}$ and {\it (ii)} the Gaussian curve on the edge of the lattice $V ( {\bm r}_{i} ) = V_{0} \left[ \exp \left( - r_{i}^{2} / 2 \sigma^{2} \right) + \exp \left( - ( r_{i} - N )^{2} / 2 \sigma^{2} \right) \right]$. Without loss of generality, we have assumed $\tilde{\Lambda}$ with the Gaussian shape located at the centre of the system, $\Lambda ( {\bm r}_{i} ) = \Lambda_{0} (t) \exp \left( - ( r_{i} - N/2 )^{2} / 2 \sigma^{2} (t) \right)$. The parameters $V_{0}$, $\Lambda_{0}$ and $\sigma$ characterize the height and the width of the  potentials, respectively.

\begin{figure}[!t]
\centering
\includegraphics[width=\linewidth]{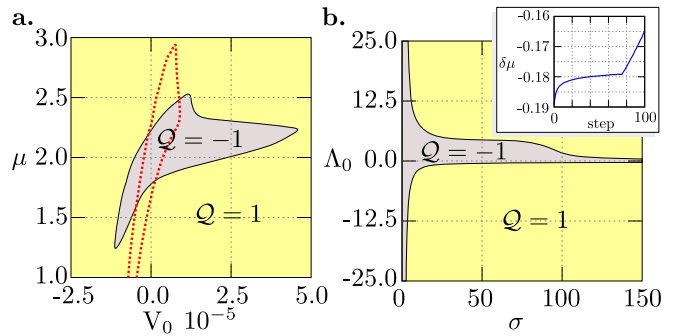}
\caption{
The phase diagram $\mu$--$V_{0}$ in the cases of (a) harmonic trap $V_{0} \delta r^{2}$ and (b) the barrier potential in the Gaussian curve form $\Lambda_{0} \exp \left( - \delta r^{2} / 2 \sigma^{2} \right)$, where $\delta r$ is the distance from the center of the system. The inset shows variation of the chemical potential $\delta \mu$ during the evolution  of the system. The (non-)trivial topological phase is displayed by (grey) yellow color.
\label{fig.charge}
}
\end{figure}

To realize a MBS in nonuniform ultracold gases, the trapping potential has to satisfy certain conditions which guarantee the appearance of a topologically non-trivial superconducting state. 
The role of the specific trapping potential and the barrier is shown in the phase diagram (Fig.~\ref{fig.charge}), where we denote the region of topological phase which supports the realization of MBS ($\mathcal{Q} = -1$) by grey color. 
For the parabolic trap (panel a), this phase exists only for the relatively flat shape (with small $V_{0}$) with a nearly constant spatial distribution of atoms.
Our results depend on the size of the system (see the dotted red line in Fig.~\ref{fig.charge}.a corresponding to a boundary of the phase with $\mathcal{Q}= -1$ for $N = 900$ sites), where the total concentration of particles $n$ is assured by appropriate $\mu$. 
We also studied the trapping potential of a Gaussian form (panel b) and found that the $\mathcal{Q} = -1$ phase exists in a broad range of the model parameters. 
We hence emphasize that profile of the trapping potential is crucial for the appearance of the MBS.
We have chosen the trapping potential of the Gaussian form with $\Lambda_{0} = 10$ and $\sigma = 20$, which gives $\mathcal{Q} = - 1$ and, as a consequence, allows to realize MBS. The numerical calculations  have been carried out for different values of $\tilde{\Lambda}$ potential. However, MBC can be induced only for these parameters, for which $\mathcal{Q} = -1$, e.g. for those shown in Fig.~\ref{fig.charge}.b.


\paragraph*{Creation of in-trap Majorana states.}

--- 
To describe the new pair of MBS induced by the additional $\tilde{\Lambda}$ potential, we have assumed time-dependent $\Lambda_{0}$ and $\sigma$. We have studied the non-Markovian dynamics, in which at every time $t$ step, the physical configuration depends only on the shape of the trapping potential and the actual barrier potential. We have investigated the evolution of the system in two steps:
($\mathcal{A}$), when the height of $\Lambda$ increases with constant $\sigma$ and 
($\mathcal{B}$), when the width of $\sigma$ increases with constant $\Lambda$.
We assumed $\Lambda (t)$ and $\sigma (t)$ given as $5 t / 75$ and $5$ in the $\mathcal{A}$ step (for $t < 75$), while $5$ and $5 + ( t - 75 ) / 4$ in the part $\mathcal{B}$ (for $t \geqslant 75$), respectively (shown by the movie POT.MP4 in the Supplementary Material).
The time $t$ corresponds to the step change of barrier $\tilde{\Lambda}$.
In both cases, the system is in topologically non-trivial state.
Upon varying the parameters of the $\tilde{\Lambda}$ potential, the topological superconducting phase can be destroyed, changing the topological invariant  $\mathcal{Q}$ from $-1$ to $1$.
To keep the fixed concentration $n$ of the fermionic atoms, we have adjusted the chemical potential. Its variation $\delta \mu$ is shown in the inset of Fig.~\ref{fig.charge}.b.

\begin{figure}[!t]
\centering
\includegraphics[width=\linewidth]{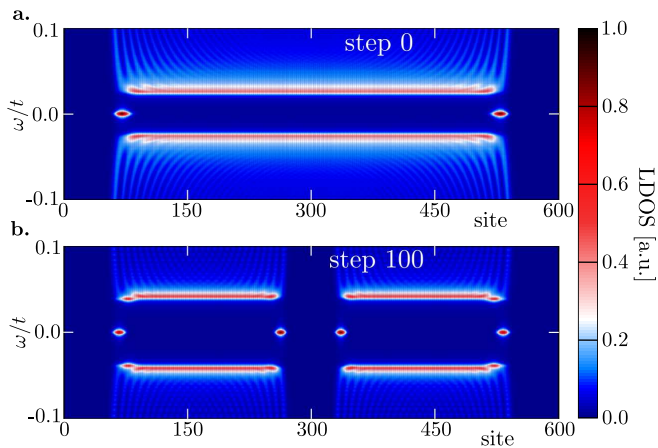}
\caption{The spatially-dependent spectrum of the fermionic atoms in the Gaussian trap at time step $0$ (panel a) and with the internal barrier, after the creation of an additional pair of MBS at time step $100$ (panel b).
\label{fig.ldos}
}
\end{figure}

For each step of the varying internal barrier, we have calculated the spatially dependent LDOS, starting from the topologically non-trivial phase with a pair of MBS (shown in Fig.~\ref{fig.ldos}.a). The evolution of $\tilde{\Lambda}$ (see the movie LDOS.MP4 in Supplemental Material) yields a new pair of MBS induced at the potential barrier. The final spectrum of the system (at time step $100$) is displayed in Fig.~\ref{fig.ldos}.b.

\begin{figure}[!b]
\centering
\includegraphics[width=0.95\linewidth]{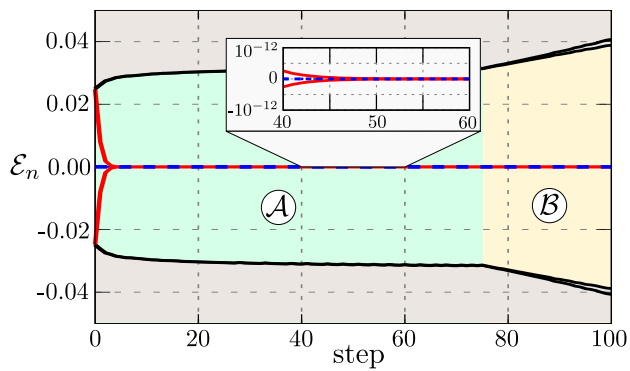}
\caption{
The low-energy eigenvalues of $\mathbb{H}$~(\ref{eq.hammatrix}) varying with the potential barrier evolution, for the parameters described in the main text. Phase $\mathcal{A}$ corresponds to  increasing height of the barrier amplitude, phase $\mathcal{B}$ to widening of the barrier, and grey denotes the region of the trivial Andreev states. Inset displays a zoom of the main figure for steps $\left\langle 40 ; 60 \right\rangle$ at which a new pair of Majorana modes is created.
\label{fig.eigen}
}
\end{figure}

\begin{figure}[!t]
\centering
\includegraphics[width=\linewidth]{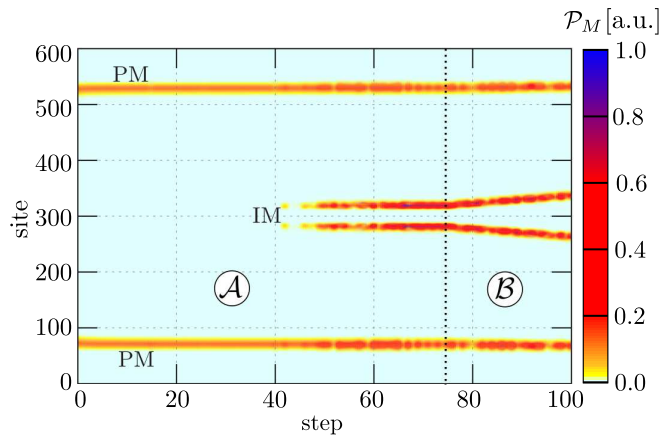}
\caption{
Density of the Majorana states $\mathcal{P}_{M}$ for the varying barrier potential $\tilde{\Lambda}$.
The primary (PM) and induced (IM) MBS are located at the edges of the trapping potential $\tilde{V}$ and near the internal barrier $\tilde{\Lambda}$, respectively. 
\label{fig.density}
}
\end{figure}

The appearance of internal zero-energy states can be observed already at step 5. By investigating the low-energy eigenvalues of the system  (Fig.~\ref{fig.eigen}), one can notice, however, that such peaks do not correspond to additional MBS. The true
Majorana modes appear around step 45 (see inset in Fig.~\ref{fig.eigen}). If such internal barrier $\tilde{\Lambda}$ was flat and spatially broad, it would induce only Andreev (non-zero energy) states~\cite{chevallier.simon.13}. This is evident from the Majorana density $\mathcal{P}_{M}$ (Fig.~\ref{fig.density}).
At each step of the potential $\tilde{\Lambda}$ evolution, the initial MBS practically does not change at all. Around step 45, we observe the emergence of a new MBS inside the 1D system, nearby the barrier $\tilde{\Lambda}$. Moreover, the change from the phase $\mathcal{A}$ to $\mathcal{B}$ does not affect much either the initial or the new MBS. The varying of the width of the barrier $\tilde{\Lambda}$ in the $\mathcal{B}$ phase only has the effect of affecting the location on the in-trap MBS. Thus, it is the $\mathcal{A}$ phase that is the crucial one for the creation of an additional MBS.
We observe that both the MBSs exist over a few sites in the 1D lattice.

 \paragraph*{Summary.}

--- We have considered a possible realization of Majorana quasiparticle states in a 1-dimensional trapped fermionic system, as proposed earlier in Ref.~\cite{jiang.kitagawa.11}. Our study clearly indicates that the MBS are very sensitive to the trapping potential, preferring its flat shapes. 
We have shown this feature by constructing a phase diagram where the topological invariant $\cal{Q}$ is plotted with respect to the parameters of the (parabolic and Gaussian) trapping potentials. In practice, their profile can be controllaby designed by the counterpropagating laser beams. 
We have also shown that the internal scattering potential is able to create an additional pair of MBS and this process is sensitive to the height of the potential. 
While the additionally created MBS is robust against changing the potential width.
This can be routinely done in ultracold gases using external laser 
beams, but similar effects can be achieved in solid state realizations of MBS (using the proximitized superconducting nanowires) by gating individual sites of the Rashba chain~\cite{aliceaoreg.11}.

Our considerations of the internal MBS in ultracold gases are appealing for a future perspective of quantum computing~\cite{rainis.12,nayak.08}.
Two pairs of MBS with opposite polarizations~\cite{sticlet.bena.12} can be written/read without any risk of decoherence, because of their topological protection. 
If the saved information is copied into separate parts in the system, then information will survive in one of the qubits, while the second qubit can be used for additional quantum operations. 
Finally, the initial information can be confronted with the results of the 
computations performed in the meantime.


\begin{acknowledgments}
We thank Krzysztof Cichy for careful reading of the manuscript, valuable comments and discussions. We also thank Ravindra Chhajlany, Jelena Klinovaja, Roman M. Lutchyn, Maciej M. Ma\'{s}ka, Pascal Simon, Jakub Tworzyd\l{}o for many fruitful discussions.
This work was supported by the National Science Centre (NCN, Poland) under grants UMO-2016/20/S/ST3/00274 (A.P.) and  DEC-2014/13/B/ST3/04451 (T.D.)
\end{acknowledgments}

\bibliography{biblio}

\end{document}